\documentclass[12pt,preprint]{aastex}

\newcommand{\be}{\begin{equation}}
\newcommand{\e}{\end{equation}}
\newcommand{\f}{\frac}

\shorttitle{CMB distortions due to IGM}

\received{/\the\month/\the\day/\the\year}
\begin{document}
\title{The Contribution of the Intergalactic Medium to Cosmic Microwave Background
Anisotropies.}

\author{F. Atrio--Barandela}
\affil{F\1sica Te\'orica, Universidad de Salamanca.\\ 
Plaza de la Merced s/n.  37008 Salamanca, Spain} 
\email{atrio@usal.es}

\and

\author{J.P. M\"ucket}
\affil{Astrophysikalisches Institut Potsdam. \\
D-14482 Potsdam, Germany.}
\email{jpmuecket@aip.de}

\begin{abstract}
We compute the power spectrum of the Cosmic Microwave Background 
temperature anisotropies generated by the Intergalactic Medium.
To estimate the electron pressure along the line of sight and its 
contribution to the Sunyaev-Zeldovich component of the CMB anisotropies, we assume 
the non-linear baryonic density contrast is well described by a log-normal distribution. 
For model parameters in agreement with observations and for an experiment operating
in the Rayleigh-Jeans regime, the largest IGM contribution corresponds to scales 
$l\approx 2000$. The amplitude is rather uncertain and  could be 
as large as $100-200\mu$K$^2$, comparable
to the contribution of galaxy clusters. The actual value
is strongly dependent on the gas polytropic index $\gamma$,
the amplitude of the matter power spectrum $\sigma_8$,
namely $C_l^{IGM}\sim(\gamma^2\sigma_8)^{12}$.
At all redshifts, the largest contribution comes from scales very close
to the baryon Jeans length. This scale is not resolved in numerical simulations
that follow the evolution of gas on cosmological scales.
The anisotropy generated by the Intergalactic Medium could make compatible the
excess of power measured by Cosmic Background Imager (CBI) 
on scales of $l\ge 2000$ with $\sigma_8=0.9$. Taking the CBI result as
an upper limit, the polytropic index can be constraint to $\gamma < 1.5$ at $2\sigma$ level 
at redshifts $z\sim 0.1-0.4$. 
With its large frequency coverage, the PLANCK satellite will
be able to measure  the secondary anisotropies coming from hot
gas. Cluster and Intergalactic Medium contributions could be
separated by cross correlating galaxy/cluster catalogs with
CMB maps. This measurement will determine the state of the
gas at low and intermediate redshifts.
\end{abstract}


\section{Introduction}

In the last decade, numerical simulations (Cen \& Ostriker 1999; Dave et al. 1999, 
2001) and observational evidence (Rauch 1998; Stocke, Shull \& Penton 2004) indicate that the highly 
ionized intergalactic gas has evolved from the initial density perturbations into a complex
network of mildly non-linear structures in the redshift interval $0<z<6$, called cosmic web.
This structure could contain most of the 
baryons in the Universe (Rauch et al. 1997; Schaye 2001; Fukugita \& Peebles, 2004). 
With cosmic evolution, the fraction 
of baryons in these structures decreases as more matter is concentrated within 
compact virialized objects. The Ly$\alpha$ 
forest absorbers at low redshifts are filaments (with low HI column densities) 
containing about 30\%  of all baryons
(Stocke et al. 2004). Hydrodynamical simulations predict that 
another large fraction of all baryons resides within mildly-nonlinear 
structures which are partly shock-confined gas filaments heated up to
temperatures of $10^5-10^7 K$, called  Warm-Hot Intergalactic
Medium (WHIM). The amount of baryons within this WHIM is estimated to reach
20 - 40 \% (Cen \& Ostriker 1999; Dave et al. 1999, 2001). 

In this article we compute the electron pressure along the line
of sight of both the ionized gas in the Ly$\alpha$ forest and the hot
gas in the WHIM. Even if the temperature of the gas in these absorbing filaments is 
relatively low (usually less than $10^5$K) the total amount of ionized gas is large. 
This electron pressure induces distortions and temperature
anisotropies on the Cosmic Microwave Background (CMB) spectrum by 
the Thermal Sunyaev-Zeldovich (TSZ) effect 
(Sunyaev \& Zeldovich, 1972, 1980) as do clusters of galaxies.
Early calculations of the TSZ power spectrum used analytical approaches
(Atrio-Barandela \& M\"ucket 1999; Komatsu \& Kitayama 1999,
Hern\'andez-Monteagudo, Atrio-Barandela \& M\"ucket 2000;
Molnar \& Birkinshaw 2000). 
Numerical simulations were soon used to make more
accurate predictions (Refregier et al. 2000, Refregier \& Teyssier 2002,
Zhang, Peng \& Wang 2002). 
Da Silva et al. (2001) studied the relative contribution from high
and low density gas and found that in all models considered, the
high density gas in halos dominated the TSZ signal.
White et al. (2002) noticed that at $l=2000$ about 25\% of the signal
came from gas in regions with density smaller than one hundred times the
cosmological mean. At $l=6000$ the contribution from the diffuse gas
was less than 2\%. 

Numerical simulations currently agree with the predictions based
on the halo model and it is reasonable to expect that the 
physics of baryons on dense environments has been well
described in cosmological simulations. What has not been proven is whether numerical
simulations have reached convergence, i.e, if further increments in resolution 
will not increase the signal (Bond et al. 2005). 
To study numerically the
evolution of the complex network of filaments that develops when the evolution of the
gas in low density regions becomes non-linear, requires computer resources
that are at present unavailable. Soft Particle Hydrodynamics (SPH) codes do not
have enough resolution and Adaptive Refinement Methods (ARM) are controlled
by density and follow the evolution of high dense regions. To study the low dense regions
with similar accuracy is computationally very expensive.

In this article we study the contribution coming from baryons
located in the low dense regions that constitute the IGM.
The web of cosmic filaments corresponds to scales
in between large scale fluctuations and strongly nonlinear scales inside
halos. The halo model is likely to fail to account for its complicated geometry
and an analytical treatment would require a different approach.
We shall use a log-normal probability distribution function (PDF) to describe
the non-linear evolution of baryons in low density environments.
Briefly, in Sec 2  we describe the model and in Sec 3 we derive the
expressions that give the contribution of the IGM to
the average y-parameter and CMB
temperature anisotropies.  In Sec. 4 we discuss our
results and their dependence with cosmological and 
physical parameters. In Sec. 5 we present our conclusions and
the observational prospects to measure the anisotropies
generated by the IGM.

\section{The Log-normal Baryon Distribution Model.}

Inverse Compton scattering of CMB  photons by hot gas along
the line of sight produces both temperature anisotropies and distortion
of the CMB spectrum. We shall assume
that baryons are distributed like a lognormal random field.
The log-normal distribution was introduced by Coles and Jones (1991)
as a model for the non-linear distribution of matter in the
Universe. Bi \& Davidsen (1997) have used it as a model to describe the
Ly$\alpha$ forest and found it reproduced the observations well.
Choudhury, Padmanabhan \& Srianand (2001) have introduced an
analytical formalism that correctly describes the clustering
properties of the neutral hydrogen in the mildly non-linear regime.
We shall follow their approach to describe the distribution
not of the neutral gas but of the highly ionized phase of
the IGM. 

If the distribution of baryons is given by a 
log-normal random field then the probability $P(\xi)$ that at any spatial
position ${\bf x}$ at redshift $z$ the baryon (non-linear) 
density contrast has the value $\xi = n_B({\bf x},z)/n_0(z)$ is given by 
\begin{equation}
P(\xi)=\frac{1}{\xi\sqrt{2\pi}\Delta_B}{\rm
 e}^{-\frac{(\log(\xi)+\Delta_B^2/2)^2}{2\Delta_B^2}}.
\label{pdf}
\end{equation}
Hereafter we shall represent the linear density contrast by
$\delta$ and the non-linear log-normal random field by $\xi$. 
The baryon number density $n_B({\bf x},z)$ is given by
\begin{equation}
n_B({\bf x},z)=n_0(z){\rm e}^{\delta_B({\bf x},z)-\Delta_B^2(z)/2},
\label{logn}
\end{equation}
where ${\bf x}$ denotes the spatial position at redshift $z$
and $|{\bf x}(z)|$ is the proper distance;
$\delta_B({\bf x},z)$ is the baryon (linear) density contrast, 
$n_0(z)=\rho_B(1+z)^3/\mu_B m_p$
and $\rho_B$, $m_p$ are the baryon density and proton mass,
respectively. 
The linear baryon power spectrum is
related to the DM power spectrum by (Fang et al. 1993)
\begin{equation}
P_B^{(3)}(k,z)=\f{P_{\rm DM}^{(3)}(k,z)}{(1+x_b^2(z)k^2)^2} ,
\label{eq:pk_b}
\end{equation}
where
\begin{equation}
x_b(z)=\f{1}{H_0}\left[\f{2 \gamma k_{\rm B} T_m(z)}
{3 \mu m_p \Omega_m (1+z)}\right]^{1/2} ,
\label{eq:xb}
\end{equation}
is the comoving Jeans length, $T_m$ is the averaged 
temperature of the IGM, $\gamma$ is the polytropic index, 
$\Omega_m$ is the cosmological fraction of matter density, 
$\mu=4/(8-5Y)$ is the mean molecular weight of the IGM and
$Y=0.24$ is the helium weight fraction.  
The Jeans length defines the scale below which 
baryon perturbations are suppressed with respect to those of the DM.
Only scales larger are allowed to grow. 
We use the log-normal statistics to describe the non-linear
evolution of those perturbations at any given epoch $z$. 
Finally,
\begin{equation}
\Delta_B^2(z)={\langle {\delta_B^2({\bf x},z)}\rangle} = 
D^2(z)\int{\f{d^3k}{(2\pi)^3}\f{P_{DM}(k)}{[1+x_b^2(z)k^2]^2}} ,
\end{equation}
where $D(z)=D(z,\Omega_{\Lambda},\Omega_m)$ is the linear growth factor.

\begin{figure}
\plotone{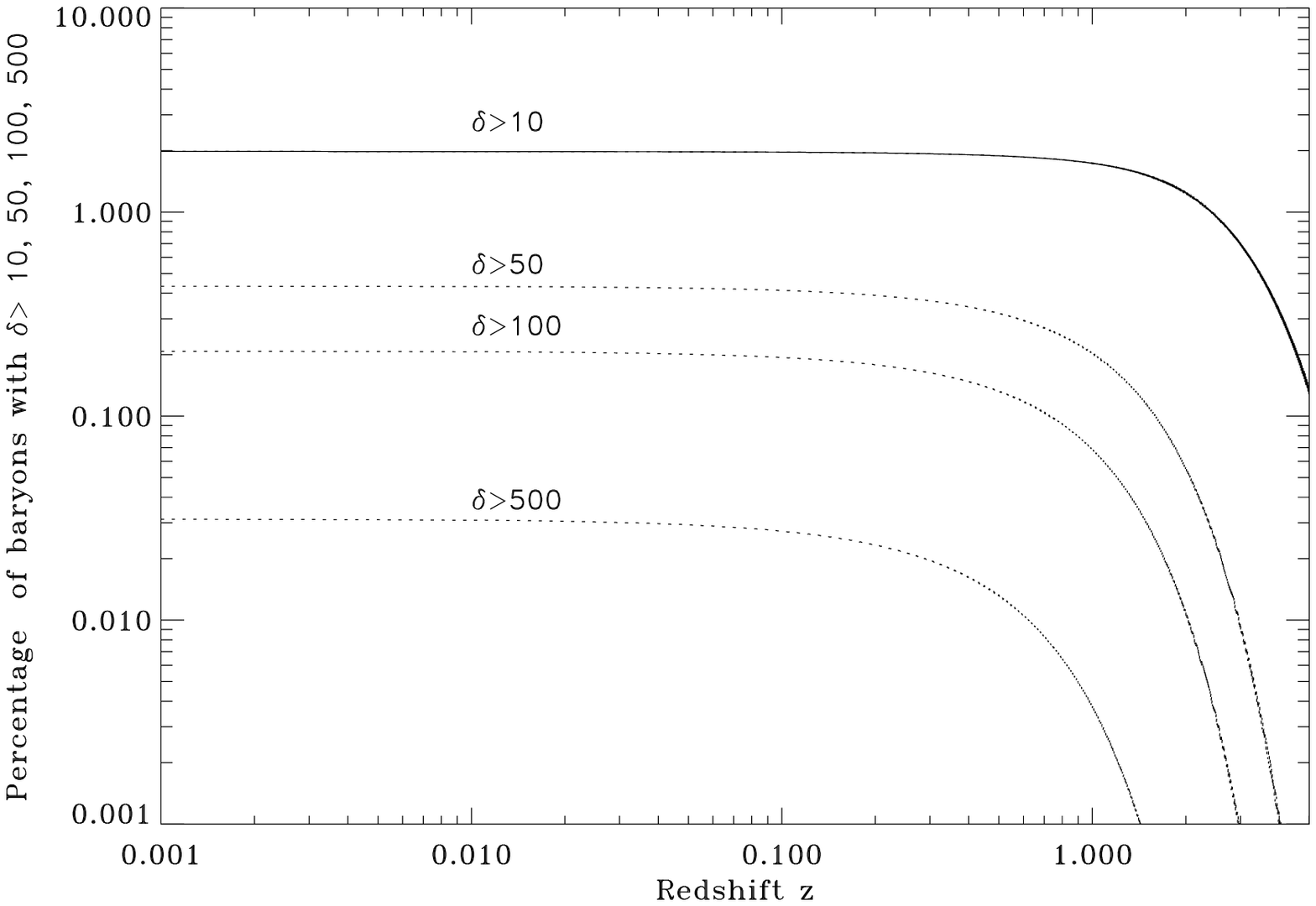}
\caption[]{
Fraction of baryons that reside in regions of density contrasts
$\xi > \delta_{max}=10, 50, 100$ and $500$ as a function of redshift.
}
\label{fig1}
\end{figure}

At each redshift, the fraction $\eta(\xi_{max})$ of matter in regions
with over-densities smaller or equal than a fixed value $\xi_{max}$ is
\begin{equation}
\eta(\xi_{max}) = 1-\int^\infty_{\xi_{max}}{P(\xi)d\xi}
\label{fraction}
\end{equation}
At each redshift, the fraction of matter in regions with over-density
larger than a fixed value $\xi_{max}$ is $1 - \eta(\xi_{max})$. 
In Fig. \ref{fig1} we plot the fraction of matter that resides in regions with 
over-density $\xi \ge 10, 50, 100$ and $500$ at different  redshifts. 
Since the log-normal PDF is
rather skewed, the fraction of matter in high dense regions is not negligible.
As the figure indicates, when the expansion of the Universe starts to accelerate
(around $z=0.5$) this fraction becomes constant. Below this redshift,
the fraction of matter with density contrasts $\xi >100$ is less than 0.3\%.

The number density of electrons $n_e$ in the IGM can be obtained by assuming ionization
equilibrium between recombination and photo-ionization and collisional ionization. At the 
conditions valid for the IGM (temperature in the range $10^4-10^7$K,
and density contrast $\xi < 100$) the degree of ionization is very high. 
Commonly $n_e = \epsilon n_B$, with $0.9<\epsilon\le 1$ depending on the 
degree of ionization. 
To compute the temperature of the IGM at each position and redshift we use a 
polytropic equation of state 
\be
T({\bf x},z)=T_0(z) \left(\f{n_B({\bf x},z)} {n_0(z)}\right)^{\gamma-1},
\e
where $T_0$ is the temperature of the IGM at mean density $n_0$
at given redshift $z$. 

\section{Comptonization Parameter and Radiation Power Spectrum.}

The contribution to the 
Comptonization parameter of a patch of hot gas of size $L$ 
at a proper distance $l=|{\bf x}|$ is (Sunyaev \& Zel'dovich 1972) 
\be
\Delta y_c(z,{\bf x}) = y_0\int_o^{L} T_e({\bf x},z)n_e({\bf x},z) dl ,
\e
where $y_0=k_B\sigma_T/m_ec^2 g(\nu)$. Constants have their usual meaning and
$g(\nu)$ is the frequency dependence of the SZ effect. The average line of 
sight contribution coming from structures located at $z$ is
\be
\Delta y_c(z) = y_0\int_o^{L} <n_e({\bf x},z)T_e({\bf x},z)> dl .
\label{yc}
\e
The average is carried out over the whole spatial volume at redshift $z$.
The total contribution up to redshift $z_f$ at which the Universe is fully re-ionized is
\be
y_{c, av}=y_0\int_0^{z_f}<n_e({\bf x},z)T_e({\bf x},z)> \f{dl}{dz}dz .
\e
Substituting the expressions of $n_e$ and $T$ obtained assuming a
lognormal gas distribution gives
\begin{equation}
y_{c, av}
= y_0\int_0^{z_f}n_0(z)T_0(z)e^{[(\gamma^2-\gamma)\Delta^2(z)/2]}\f{dl}{dz}dz .
\label{yaverage}
\end{equation}
Numerically, we shall restrict the average on eq.~(\ref{yc}) to baryons
residing in over-densities $\xi \le \xi_{max}$ (which for concreteness we shall
take equal to $100$) to exclude those baryons
that could not be correctly described by the log-normal model (see below).
Therefore, at at each redshift we compute
\be
<f(\xi)>_{|\xi_{max}} = \int^{\xi_{max}}f(\xi)P(\xi)d\xi
\label{coorect}
\e
The power spectrum contribution
of the CMB temperature anisotropies induced by the IGM can be obtained from
the 2-point correlation function of the spatial variations of the electron
pressure:
\be
C(\theta)= \int_0^{z_f}\int_0^{z_f}
\Delta y_c(z)B(\theta,z,z')\Delta y_c(z') \f{dl}{dz}\f{dl'}{dz'}dz dz' .
\label{cfull}
\e
In this expression $\Delta y_c(z)$ is given by eq.~(\ref{yc}),
$ B(\theta,z,z')=e^{[\gamma^2  Q(|{\bf x}-{\bf x'}|,z,z')]}-1$
is the normalized two-point correlation function
with
\begin{equation}
Q(|{\bf x}-{\bf x'}|,z,z')={D(z,\Omega_\Lambda,\Omega_m)D(z',\Omega_\Lambda,\Omega_m)\over
  2\pi^2}
\int^\infty_0{{P_{DM}(k)k^2\over [1 + x_b^2(z)k^2][1 + x_b^2(z')k^2]}{\sin
    (k|{\bf x}-{\bf x'}|)\over k|{\bf x}-{\bf x'}|}} . 
\label{Q}
\end{equation}
In here $|{\bf x}-{\bf x'}|$ denotes the proper distance between two patches at
positions ${\bf x}(z)$ and ${\bf x'}(z')$ separated by the angle $\theta$.
In the flat sky approximation
\be
|{\bf x}-{\bf x'}|\approx \sqrt{l_\bot(\theta,z)^2+ [r(z)-r(z')]^2} ,
\e  
where $l_\bot(\theta,z)$ is the transversal distance of two points
located at the same redshift. 
Within this approximation, the correlation function is dominated by 
patches that are physically very close.  For small 
$\theta$, the correlation between patches at different redshifts is negligible 
and $B(\theta,z,z')\simeq B(\theta,z,z)\delta_{Dirac}(z-z')$ is accurate at the 1\% level.
Eq.~(\ref{cfull}) can be simplified to give
\be
C(\theta)=y_0^2\int_0^{z_l}{\left[\f{dl}{dz}\right]^2
n^2_0(z)T^2_0(z)e^{\gamma(\gamma-1)\Delta^2(z)}[e^{\gamma^2Q(\theta,z)}-1]}dz
\label{ctheta}
\e
This approximation to the eq.~(\ref{cfull})
fails at large angular scales but, since at those scales the correlation is negligible,
it does not affect the numerical results while greatly speeds up the computer
code. Like  for the Comptonization parameter $y_{c,av}$, we restrict the average to
baryons within the mildly-nonlinear regime, i.e., $\xi \le \xi_{max}=100$.
Finally, the power spectrum can be obtained by Fourier transform:
\be
C_l^{IGM} = 2\pi\int_{-1}^{+1}{C(\theta)P_l(\cos\theta)}d(\cos\theta) ,
\label{cl}
\e
where $P_l$ denotes the Legendre polynomial of multipole $l$.

Since $y_{c,av}$ and $C_l^{IGM}$ depend on the electron
pressure and not separately on the IGM temperature or density,
eqs.~(\ref{yaverage}) and (\ref{ctheta}) scale with IGM mean temperature $T_0$ and ionization
fraction $\epsilon$ as $(T_0\epsilon)$ to some power. 
The uncertainty in the degree of gas ionization is smaller than the one on the
temperature $T_0$ at mean density, so we shall not consider it any further. 
In our numerical results we shall take $\epsilon = 1$.

\section{Numerical Results.}

To compute the contribution of the IGM to CMB temperature anisotropies,
we take the concordance model as our fiducial cosmological model:
$\Omega_\Lambda=0.73$, $\Omega_{DM}=0.23$, $\Omega_{B}=0.04$, $h=0.71$ and
$\sigma_8=0.9$, in agreement with WMAP results (Spergel et al. 2003). 
Except when specified otherwise, we present our results for $g(\nu)=1$. 
The physical parameters describing IGM thermal evolution
history: the temperature $T_0$ at mean density, the Jeans length -fixed
by $T_m$- and the polytropic index $\gamma$, are the free parameters of our model. 
When the temperature of the IGM increases (e.g., during the
re-ionization of He at $z\approx 3.$) so does the Jeans length.
Then, perturbations that were previously evolving are frozen or partly damped.
Suppression of power on smaller scales
can also happen at an early epoch by energy injection (Springel et al. 2001).
Since the contribution of those scales 
will be erased, one would require a detailed study of the evolution 
of $T_m$ with redshift to estimate the CMB temperature anisotropies.  
To be conservative, we shall take it equal to the largest
admissible value instead of the average IGM temperature. 
The HeII re-ionization at $z\approx 3$ requires temperatures 
larger than $5\times 10^4$K (Schaye et al. 2000). 
Hui \& Haiman (2003) argued that the IGM reached higher values during 
its thermal history. Analyzing the SDSS data, Viel \& Haehnelt (2005) found 
that the temperature range is weakly constrained and gave an upper 
bound of  $T_m\approx 2\times 10^5$K.  
For our numerical estimates we adopt a maximum value of $T_m =1\times 10^5$K.
With respect to the temperature at mean density we take an average of
$T_0 = 1.4\times 10^4$K according to the lower values obtained by Hui \& Haiman (2003). 
This temperature is mainly determined by the equilibrium of the 
photoionization due to the UV background radiation and recombination of 
hydrogen at mean density. This value is also within the range obtained
from the analysis of the QSO absorption lines (Stocke et al.  2004). 
Finally, our numerical estimates depend critically on the redshift evolution of 
$\gamma$, that is very uncertain and strongly model dependent. Since our results show 
(see below) that most of the contribution
to the CMB temperature anisotropy is generated in the redshift interval
$z\approx 0.1-0.4$, we can fix $\gamma$ to be the average value in
that redshift interval. 

Our current ideas of galaxy formation suggest 
that above $z=6$ an increasing fraction 
of the gas is neutral and with low temperature; on those grounds,  
we do not expect a large contribution from earlier epochs.
Therefore, all integrations were carried up to $z_f=6$,
the epoch when the re-ionization can be considered complete.
 
\subsection{Mean Comptonization.}
\begin{figure}
\plotone{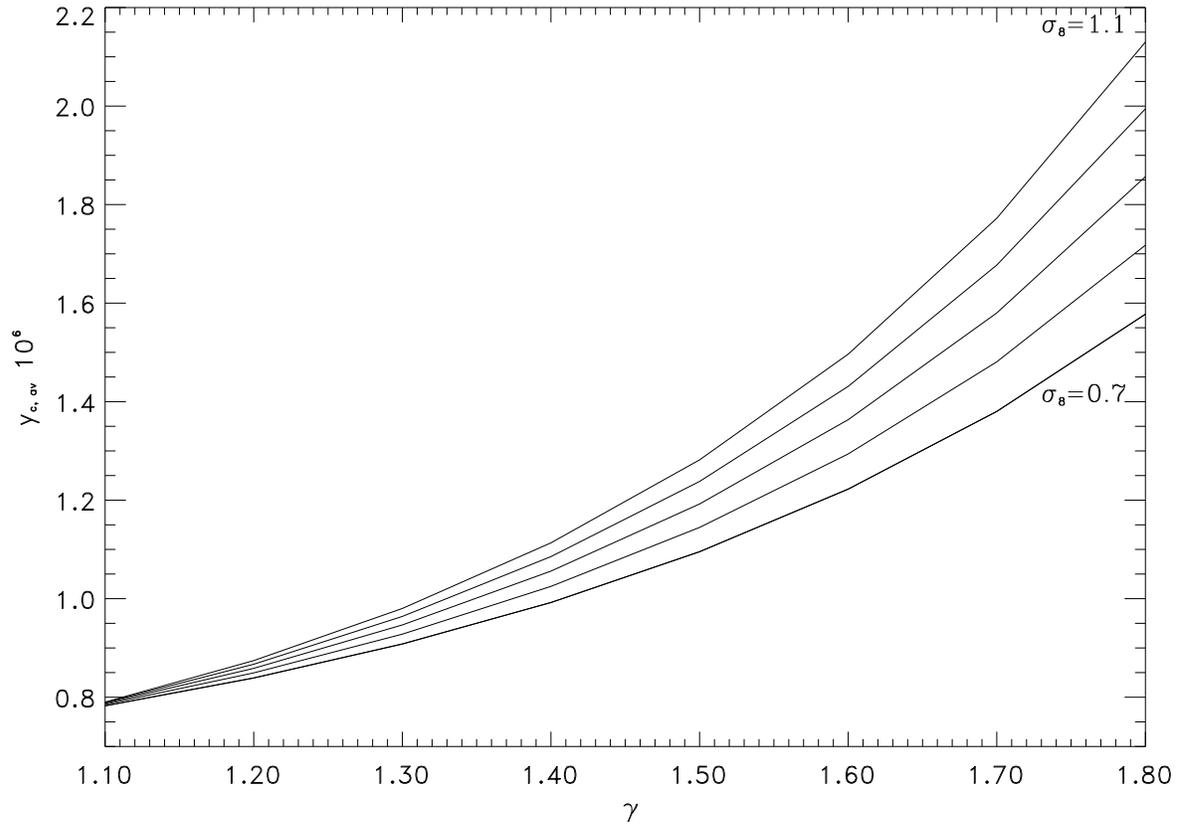}
\caption[]{Mean Comptonization parameter $y_{c, av}$ as a function of 
the polytropic index $\gamma$. Curves correspond to different value of $\sigma_8$.
From top to bottom, curves decrease in units of one tenth.
}
\label{fig2}
\end{figure}

Eq.~(\ref{yaverage}) gives the average $y$-parameter distortion produced by
the IGM.  In Fig. \ref{fig2} we show the dependence of the amplitude 
of the average Comptonization parameter $y_{c, av}$ with respect to the parameter $\gamma$.
In the figure, $\sigma_8$ varies from 0.7 (bottom) to 1.1 (top) 
in units of one tenth.

\subsection{Radiation Power Spectrum.}
\begin{figure}
\plotone{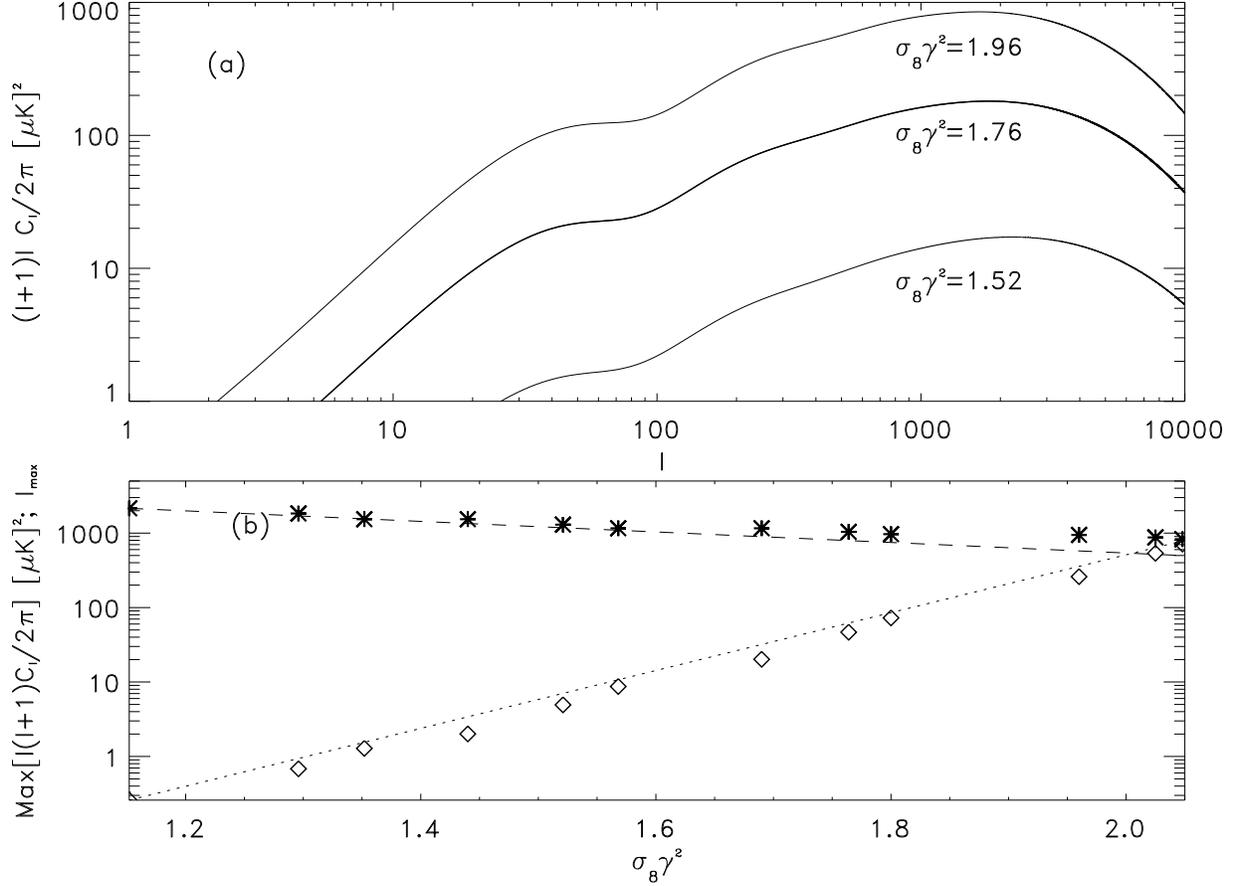}
\caption[]{(a) CMB radiation power spectrum. 
The middle curve corresponds to $\gamma = 1.4$, $\sigma_8 = 0.9$. 
(b) Best fit to the amplitude (dotted line) and location
(dashed line) of the radiation power spectrum maxima as a function of combined
gas and cosmological parameters ($\sigma_8\gamma^2$). 
Asterisk and diamonds correspond
to the models actually computed. The y-axis gives the maximum
value in ($\mu$K$^2$) -diamonds- and the multipole $l_{max}$ corresponding to
the maxima -asterisks-.}
\label{fig3}
\end{figure}

In Fig. \ref{fig3}a we show several spectra for different values of
$\sigma_8\gamma^2$. The amplitude of the radiation power spectrum at all scales is strongly
dependent on this product (see eqs.~\ref{cfull}-\ref{ctheta}). 
The small wiggle at $l=10-50$ is caused by using the approximated correlation function
given by eq.~(\ref{ctheta}) at large angular scales. 
In Fig. \ref{fig3}b, diamonds show the variation
of the power spectrum maximum amplitude as a function of $\sigma_8\gamma^2$. 
It corresponds to a scaling $C_{l,max}^{IGM}\sim (\sigma_8\gamma^2)^{12}$.
Due to this strong dependence, if $\sigma_8$ is known, even an order 
of magnitude estimate of the  
IGM contribution to temperature anisotropies will give a rather accurate
measurement of the polytropic index $\gamma$. 
In the same Fig. \ref{fig3}b, asterisks show the location of the 
radiation power spectrum maxima and the dashed line corresponds to the best fit.
Here the dependence is much weaker and,  in our range of cosmological
parameters, the maximum anisotropy corresponds to angular scales ranging from 5 to 10 arcmin.

\subsection{Contribution of different redshift intervals.}
\begin{figure}
\plotone{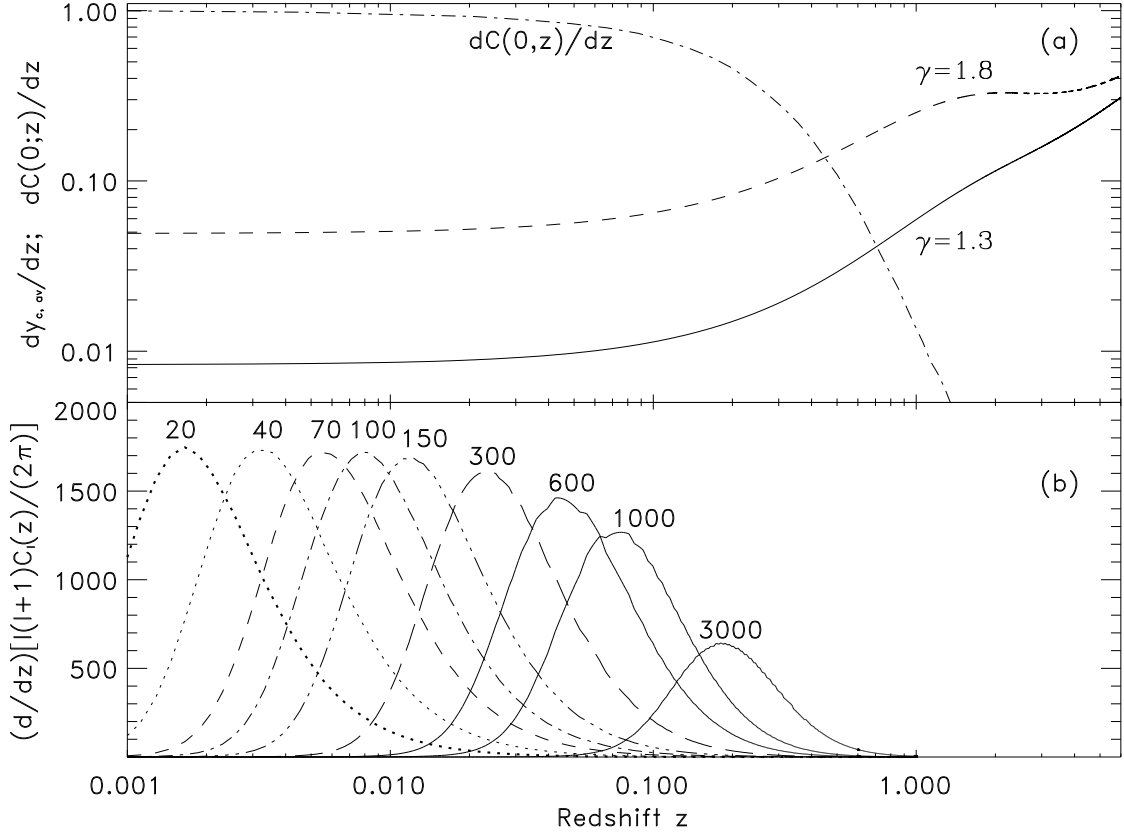}
\caption[]{(a) Contribution of the IGM at different redshift intervals,
to the Comptonization parameter in redshift bins of $\Delta z = 0.001$, 
for different parameters. 
The dash-dotted line shows the time dependence of $dC(0;z)/dz$ 
(normalized to unity at $z=0$).
(b) Contribution to the radiation power spectrum of redshift bins
of width $\Delta z = 0.001$ as a function of redshift $d[l(l+1)C_l]/dz$ 
for fixed multipoles $l$ ($\gamma=1.4, \sigma_8=0.9$). }
\label{fig4}
\end{figure}

Due to its definition, $T_0$ is expected to vary little with redshift, and
any time dependence can be easily incorporated into the analytical expressions.
On the other hand, the polytropic index $\gamma$ is strongly dependent on 
the thermal history of the IGM. 
In our calculations we have assumed that $T_0$ and $\gamma$ are strictly constant
during the cosmic evolution of the IGM from re-ionization till today. Even in this simplified model
not all redshifts contribute equally to CMB distortions and temperature anisotropies.
In Fig. \ref{fig4}a we show the differential growth of the Comptonization parameter
with respect to redshift $dy_{c, av}/dz$ 
for different $\gamma$. In the figure, the redshift intervals are
$\Delta z = 0.001$. Let us remark that when $\gamma$ is 
small, the contribution of the IGM to $y_{c, av}$ at high redshifts ($z> 0.3$) 
is much higher than at smaller redshifts.   
For large values of $\gamma$, the gas located 
at redshifts $z>1$ gives approximately equal contributions and even the low-redshift gas 
contributes significantly. In all cases, the contribution
increases close to $z=6$ and the choice of the upper limit of integration can affect 
the average distortion. As far as the IGM is well described by 
our model out to $z_f=6$ our results must be taken as lower limits. 
For comparison, we also show the contribution of the different redshift
intervals to the correlation function: $dC(0,z)/dz$ for $\gamma = 1.4$, 
normalized to unity at $z=0$.
In Fig. \ref{fig4}b we show the differential redshift 
contribution to the radiation power 
spectrum: $d/dz[l(l+1)C_l^{IGM}/2\pi]$, for different multipoles. 
There is a clear difference with the behavior shown in Fig. \ref{fig4}a:
at each multipole the signal comes preferentially from a narrow redshift range. 
Even if the maximum value decreases for increasing $l$, 
the effective width of redshift intervals dominating the contribution
increases and the overall maximum in the full spectrum occurs at $l \approx 1000 - 3000$,
as indicated by Fig. \ref{fig3}a.

In Figs. \ref{fig2} and \ref{fig3} we are implicitly assuming that 
all baryons are in the IGM (see eq.~(\ref{logn})). 
At high redshifts ($z =2 - 4$) the entire baryon content of the universe
can be accommodated within the warm ($\sim 10^4$~K) photo-ionized 
IGM. At low redshifts, 
the combined fraction of baryons in warm photo-ionized IGM together
with those in the WHIM could be as large as 70-80\%.
As indicated in Fig. \ref{fig4}b the radiation power spectrum originates on 
a very narrow redshift range,
and if during that period the fraction $f$ of baryons in the IGM is kept constant
at, say, 70\%, the amplitude of the power spectra, that scales like $f^2$, would be reduced 
by a factor 2. The effect would be very small for the y-parameter
since first $y_{c,av}$ scales linearly with f, and second the gas at higher
redshifts, where $f\approx 1$, also contributes. 
Considering the strong dependence of $C_l^{IGM}$ with $\gamma$ and $\sigma_8$,
the effect of this uncertainty in the upper limit derived above is not significative.

\subsection{Contribution of Different Scales.}
\begin{figure}
\plotone{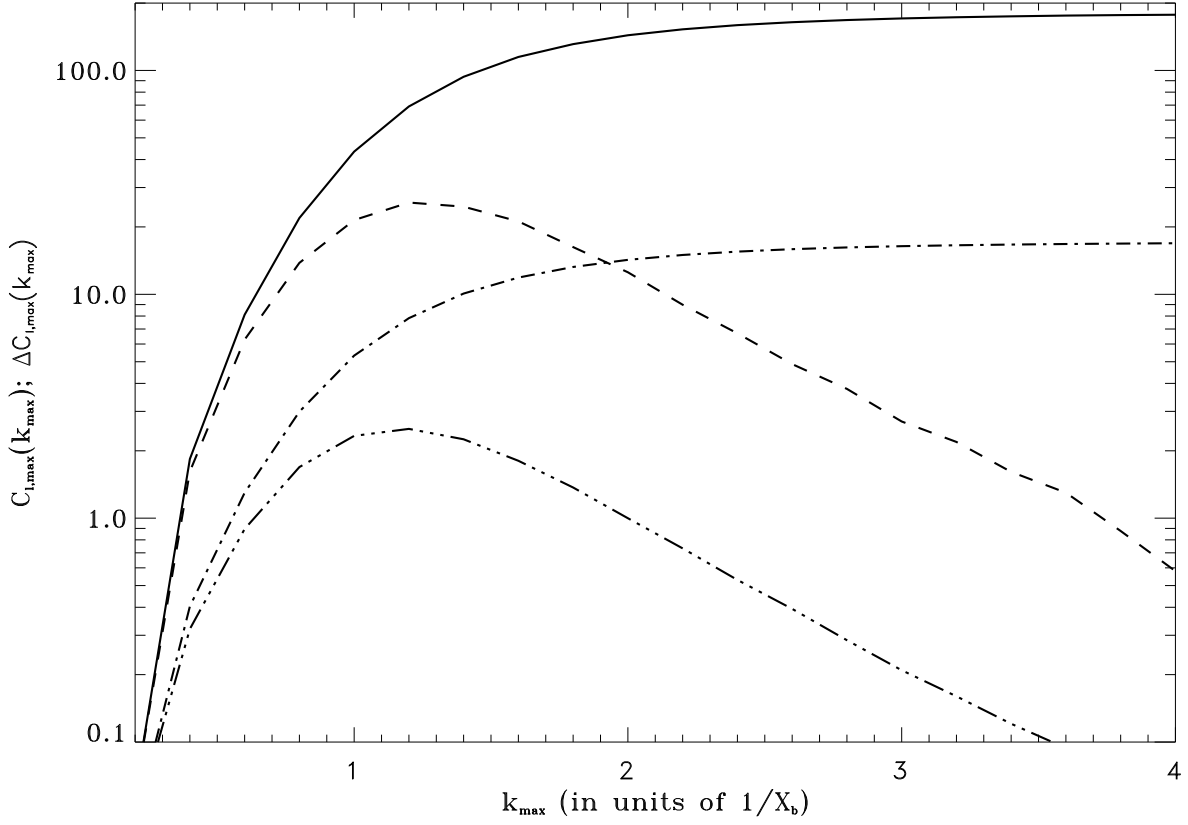}
\caption[]{
Upper curves: contribution to $C_{l,max}$ of all scales $k\le k_{max}$ 
($k_{max}$ in units of $x_b$), i.e.,
when the integral over $k$ in eq.~(\ref{Q}) is restricted to that interval
Solid line corresponds to $\gamma=1.4, T_m=10^5$K,
dash-dotted line to $\gamma=1.3$K.
Lower curves (dashed and dash-triple dotted) : contribution per bins of width $\Delta k = 0.2/x_b$
to $C_{l,max}$.}
\label{fig5}
\end{figure}

If the main drawback of analytical treatments is to be based on simplifying assumptions 
and to require scaling relations derived from observations, current numerical simulations
lack spatial resolution in large enough volumes to resolve the complex
gas dynamics (Bond et al. 2005). 
Our results, detailed above, seemingly contradict those of numerical
simulations carried out until present. For example,
White, Hernquist \& Springel (2002) and da Silva 
et al. (2001)  analyzed whether the majority of the contribution to the 
TSZ angular power spectrum came from diffuse gas or gas within haloes.
Fig. \ref{fig5} is useful  to understand the different outcome between our 
analytical estimates and the results obtained using cosmological hydro-simulations. 
In the figure, we plot the contribution of the
different scales to the radiation power spectrum. We particularize for
$l\approx 2000$ that in all models is close to the largest amplitude of the
radiation power spectrum $C_l$. In the upper two curves we integrate eq.~(\ref{Q}) from $k=0$
to $k_{max}$, expressed in units of the inverse Jeans length $1/x_b$.
The solid line corresponds to $\gamma=1.4, T_m=1\times
10^5$K and the dashed line to $\gamma=1.3$ at the same $T_m$.
Let us remark that even for a fixed $T_m$ the Jeans length varies with
the polytropic index $\gamma$.  The lower two 
curves show the differential contribution to $C_{2000}$ of
different scale intervals. We integrate eq.~(\ref{Q}) in bins of
width $\Delta k=0.2/x_b$. Lines correspond to the same parameters as before.
The amplitudes are different depending on model parameters, but
both curves have very similar shape.
The plot clearly shows that the main contribution to $C_l$ comes from
scales in the range $kx_b\simeq (0.5 - 3)$. About 80 to 90\% of the total
power comes from scales in that range, the exact figure
depending on model parameters.

In a model with $\gamma=1.4$, $T_m=1\times 10^5$K, the Jeans length
is $x_b= 470 h^{-1}$kpc. 
High-resolution hydrodynamical simulations with 
ARM techniques are extremely good in dealing with the gas behavior within 
high density regions since the refinement is governed by the local density. 
For example, Refregier \& Teyssier (2002)
studied the evolution of DM and gas with resolutions
of 96$h^{-1}$kpc at $z\sim 5$ to 12$h^{-1}$kpc at $z=0$
in the densest regions using an adaptive mesh refinement algorithm.
But their resolution was much smaller in low density regions, where
most of the baryons reside.
SPH do not yet account for contributions that comes from  scales 
$k\ge 2/x_b$ or are just on the edge of the necessary resolution.
A SPH code evolving $2\times (216)^3$ particles in a box of $100h^{-1}$Mpc
and cell size of $370 h^{-1}$kpc (i.e. White et al. 2002) 
could reach enough spatial resolution to follow the gas dynamics on 
scales close to the Jeans length, this is not the case.
Spatial resolution is not cell size but the minimun 
scale below which the code is not able to solve the evolution equations. Particle
mesh codes include a force cut-off at small distances resulting on a
damping of the power spectrum up to 8 times the cell size (Refregier \& Teyssier 2002).
Even if  White et al. (2002) had a effective resolution 
much larger than their cell size and could not account for the effect
of gas on scales close to the Jeans length, they did find a 25\% contribution
coming from regions with density contrast $\le 100$ at $l=2000$. This contribution was 
less than 2\% at $l=6000$ much as it could be expected if
the signal was due to diffuse gas. As Fig. \ref{fig3}a shows, between those multipoles
the power decreases by a factor $3-10$, depending on the IGM temperature
and polytropic index. 
 
\subsection{Contribution of different density contrasts and Jeans length.}
\begin{figure}
\plotone{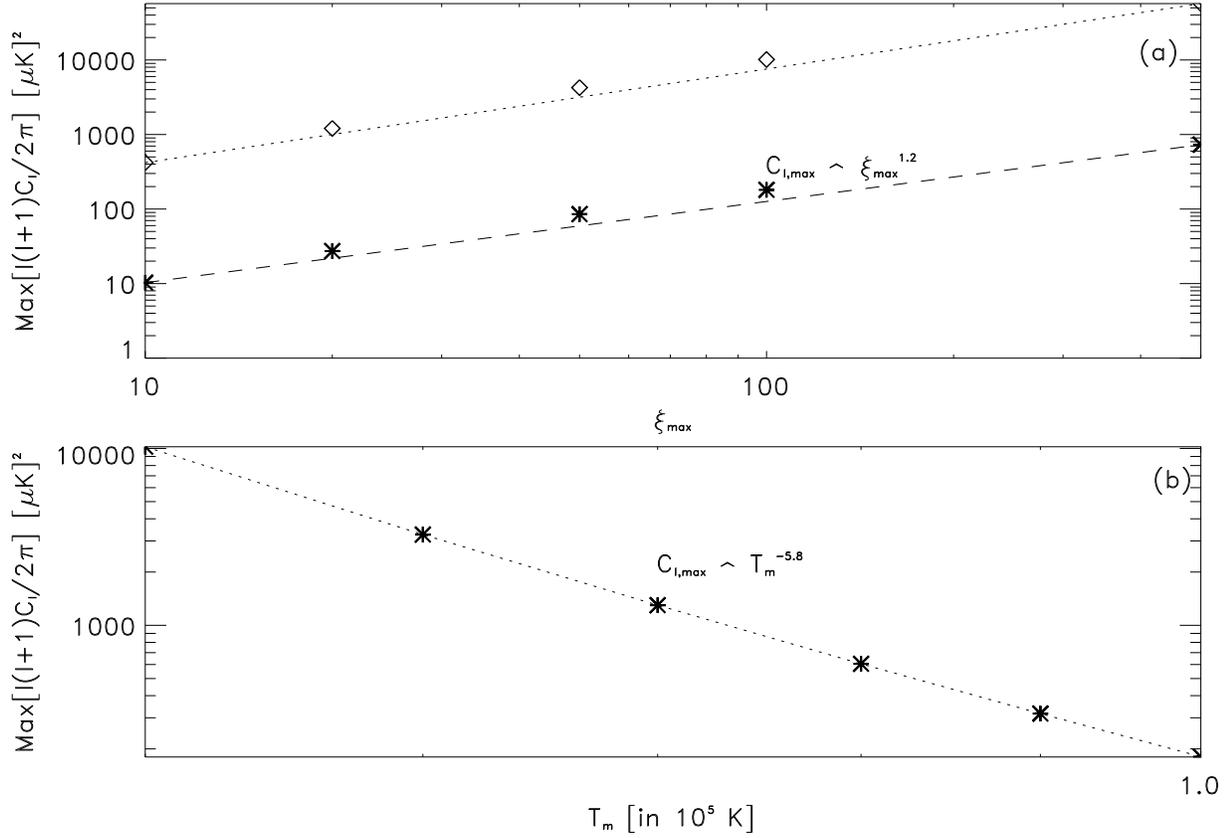}
\caption[]{Scaling of the maximun amplitude of the
radiation power spectrum $C_{l,max}$ 
 with (a) $\xi_{max}$ and (b) $T_m$. All models have $\sigma_8=0.9$, $\gamma =1.4$.
In (a) dotted line corresponds to $T_m=5\times 10^4$ K and dashed line to $T_m = 1\times 10^5$ K.
In (b) $\xi_{max}=100$.}
\label{fig6}
\end{figure}

Analytic calculations based on the halo model correctly
account for the SZ contribution of gas in clusters of galaxies
and collapsed objects. 
On the contrary, a log-normal description of baryons in the IGM is only valid for a limited
range of over-densities. In the redshift interval [0.1,0.4], where most
of the temperature anisotropy is generated,  the fraction of matter with 
$\xi >100$ is less than 0.3\% (see Fig. \ref{fig1}). Since the log-normal PDF weights heavily
the high density regions, even this fraction could have a large contribution.
Observationally, it has been established
that the model describes rather well Ly-$\alpha$ clouds
with over-densities $\xi\le 50$ by $z=3$. 
Recent analysis, see e.g.,  Tatekawa (2005), Kayo, Taraya \& Syto (2001),
have shown that as a result of cosmic evolution,
the density distribution becomes increasingly better described by a log-normal
PDF at $z\to 0$. In particular, it was shown by Kayo et al. (2001) that it accurately 
describes the density distribution even in the non-linear regime up to $\xi<100$. 
Baryons at much higher densities ($\xi>500$)
will cluster in halos and experience different
astrophysical processes such as shock heating, radiative effects,
star formation, energy injection through supernovae
explosions, etc., occurring on galaxy and cluster scales.
Those baryons can not be described as a gas with uniform temperature
and polytropic equation of state. As remarked in Sec. 3,
averages in eqs.~(\ref{yc}) and (\ref{cfull}) have been carried out excluding 
regions with over-densities $\xi>100$.
In Fig. \ref{fig6}a we examine how the maximum amplitude of the radiation
power spectrum $C_{l,max}$ scales with $\xi_{max}$ for two different values of $T_m$. 
The scaling is rather weak compared to the dependence on the polytropic index
$\gamma$.

The Jeans length measures the minimum scale that is gravitationally unstable.
It varies with redshift as different physical processes heat
and cool the gas. Then, the evolution of scales close to the Jeans length becomes
rather complex. We simplify their treatment assuming $T_m$ to be fixed 
and equal to its maximum value throughout its thermal evolution.
Since the baryon power spectrum in eq.~(\ref{eq:pk_b}) is damped on scales
smaller than the Jeans length, this assumption reduces the contribution of
those intermediate scales.  In Fig. \ref{fig6}b
we plot the dependence of the maximum amplitude of $C_{l,max}$ with $T_m$.
As expected, larger $T_m$ lead to a smaller contribution since a smaller
number of scales is included.
To summarize, the scaling behavior represented in 
Figs. \ref{fig3}b and \ref{fig6} is 
\begin{equation}
C_{l,max}\sim \sigma_8^{12}\gamma^{24}T_m^{-6}\xi_{max}^{1.2}
\end{equation}
For comparison, $C_l^{clusters}\sim \sigma_8^{6}$
obtained by Komatsu \& Kitayama (1999) from analytical estimates.
Small variations on $\gamma$ can produce a strong change in the
amplitude of the radiation power spectrum. Even 
if $T_m$ is only known up to an order of magnitude and our model can only be applied
up to over-densities $\xi \le 20$, we could still obtain strong constraints on 
$\gamma$ from observations of CMB temperature anisotropies.

\subsection{IGM contribution to CBI scales.}

\begin{figure}
\plotone{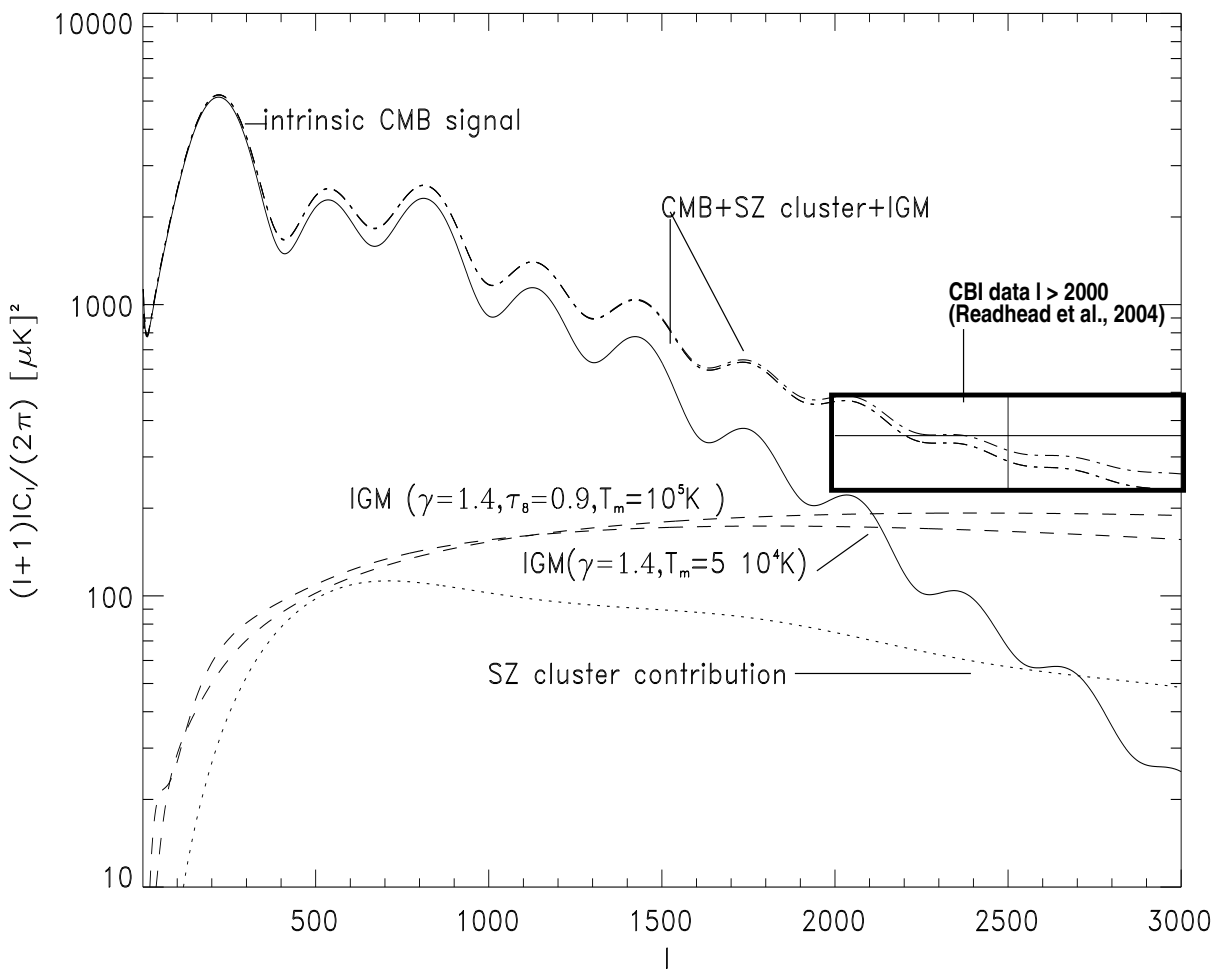}
\caption[]{TSZ radiation power spectrum component from clusters (dotted line)
and IGM (dashed) for two different polytropic indeces, 
intrinsic CMB temperature anisotropies (solid) and the sum of the three components
(dot-dashed). The TSZ power spectrum has been rescaled to 32 GHz.
The box gives the CBI data at the scales of interest.
}
\label{fig7}
\end{figure}

In Fig. \ref{fig7} we compare the CMB temperature anisotropy power spectrum 
of cosmological origin (solid line) with TSZ contribution of clusters 
of galaxies (dotted line) and the IGM (dashed line) with 
$\gamma = 1.4, T_m=10^5 K$ and $\gamma = 1.27, T_m=5\times 10^4$.
We adopted $g(32GHz)=-1.96$ to rescale the TSZ power spectra to the operating frequency of 
the Cosmic Background Imager (CBI) experiment (Readhead et al., 2004).  
CBI showed an excess over the cosmological radiation power spectrum  at $l>2000$.
The result and its 1$\sigma$ error box is included in the figure.
Depending on model parameters, TSZ cluster and IGM  components could have similar 
or very different amplitudes and shapes. Sadeh \& Rephaeli (2004)
have demonstrated that the radiation power spectrum generated by clusters depends
on the assumed mass-temperature relation and gas evolution.
The cluster power spectrum represented in Fig. \ref{fig7} has been generated
using an analytical model that does not peak at $l=2000$.
Bond et al. (2005) found that to explain the CBI excess 
with the anisotropy generated by clusters requires $\sigma_8=1.0$ or larger. 
Taking into account the contribution of the IGM, it can be seen in Fig. \ref{fig7} that
if $\sigma_8=0.9$ and $T_m$ and $\gamma$ vary
within the intervals $5\times 10^4 \le T_m \le 10^5 K$ and $1.27 \le \gamma \le 1.4$, 
the combined power spectrum CMB+SZ cluster+IGM is fully consistent
with the CBI observations.  Turning the argument around,  if the CBI data 
is assumed to be an upper limit to the total anisotropy on scales
$l> 2000$, then $\gamma > 1.5$ is ruled out at the $2\sigma$ level. 
Since the contribution to those scales comes mostly from $z=0.1-0.4$ (see Fig. \ref{fig4}b),
this upper limit applies to the gas at that redshift.

\section{Discussion.}

In this article we have shown that ionized gas, in the deep potential wells
of clusters of galaxies and in the IGM, have a significative effect on the
CMB, generating both temperature anisotropies and spectral distortions. 
While the latter is proportional to the
electron pressure along the line of sight, the former depends
on the clustering properties of the hot  gas  (Hern\'andez-Monteagudo et al. 2000).
The high amplitude
of the radiation power spectrum, similar in magnitude to the contribution of clusters,
is due to the non-linear evolution and high degree of clustering of the baryonic
matter at low redshift.  
It is the lognormal distribution of the ionized gas
that drives the strong dependence of the amplitude of the radiation
power spectrum with $\gamma$ and $\sigma_8$.  
This extra component can make compatible the
CBI measured power excess at $l\sim 2000$ with $\sigma_8=0.9$.
At all redshifts, the largest contribution comes from scales
around the baryon Jeans length. In cosmological simulations set to
compute the TSZ contribution to the Cosmic Microwave Background temperature
anisotropies, this scale is not well resolved, what explains why this contribution
has not yet been identified in numerical simulations. 

The log-normal model is remarkable for predicting a strong dependence of the radiation
power spectrum with two parameters: $C_{l,max}\sim (\gamma^2\sigma_8)^{12}$.
Even if the range of scales, densities and redshifts 
to which the model can be applied is rather uncertain,
those parameters produce minimal variations in shape of the
radiation power spectrum and the variations in the amplitude
are much smaller than that of $\gamma$ and $\sigma_8$. 
Remarkably, the anisotropy due to the IGM is generated
in a narrow redshift interval.  Assuming 
$\sigma_8=0.9$, one can obtain a firm upper limit of $\gamma \le 1.5$ at
the 2$\sigma$ confidence level in the redshift range [0.1-0.4].
As different redshift intervals dominate the anisotropy at different angular scales,
measurements of the power spectrum at different $l$ will allow to determine
the polytropic index $\gamma$ and the state of the ionized gas at different redshifts.

By hypothesis, the gas obeys a polytropic equation of state. This cannot
account for the effect of the shock heated gas of the WHIM leading to high
temperatures ($T \approx 10^6 - 10^7 K$). Thus, this baryon component might
provide an 
extra TSZ contribution at redshifts close to zero. 
To distinguish the IGM TSZ effect from the one coming from
clusters of galaxies, it is necessary to use their different statistical properties. 
Since the TSZ effect is independent of redshift, the cluster signal will correlate 
with cluster positions on the sky. Cross-correlation of cluster catalogs
with CMB maps  opens the possibility of determining the cluster contribution
and to separate the IGM component. Hern\'andez-Monteagudo, G\'enova-Santos
\& Atrio-Barandela (2004) and Afshordi, Lin \& Sanderson (2005)
have recently carried out such an analysis using WMAP data.
They found strong evidence (at the 5 and 8$\sigma$  levels, respectively)
of a TSZ contribution to the radiation power spectrum
due to clusters but the data was not sensitive
enough to yield the radiation power spectrum.
The PLANCK satellite, with its large frequency coverage, will be well suited
for measuring the TSZ power spectrum. It is worth to explore correlation
techniques that will permit to separate the cluster from the IGM component.
Measurements of the IGM power spectrum at different
multipoles will provide a measurement
of the state of the gas (temperature and polytropic index) at different
redshifts.

\acknowledgments

JPM  thanks the KITP for financial support. This research was supported in part 
by the National Science Foundation under Grant No. PHY99-0794.
F.A.B. acknowledges
financial support from the Spanish Ministerio de Educaci\'on y Ciencia
(projects BFM2000-1322 and AYA2000-2465-E) and from the Junta de
Castilla y Le\'on (projects SA002/03 and SA010C05).

\clearpage

\clearpage

\clearpage

\clearpage

\clearpage

\clearpage

\clearpage

\end{document}